# Si/AlN p-n heterojunction interfaced with ultrathin SiO$_2$

Haris Naeem Abbasi[1, †], Yi Lu[1, †], Jie Zhou[1], Ding Wang[2], Kai Sun[3], Ping Wang[2], Jiarui Gong[1], Dong Liu[1] Yang Liu[1], Ranveer Singh[1], Zetian Mi[2], and Zhenqiang Ma[1, *]

[1]*Department of Electrical and Computer Engineering, University of Wisconsin-Madison, Madison, Wisconsin, 53706, USA*

[2]*Department of Electrical Engineering and Computer Science, University of Michigan, Ann Arbor, Michigan, 48109, USA*

[3]*Department of Material Science and Engineering, University of Michigan, Ann Arbor, 48109, USA*

†These authors contributed equally.

*Author to whom correspondence should be addressed. Email: mazq@engr.wisc.edu, ztmi@umich.edu

# Abstract

Ultra-wide bandgap (UWBG) materials hold immense potential for high-power RF electronics and deep ultraviolet photonics. Among these, $Al_xGa_{1-x}N$ emerges as a promising candidate, offering a tunable bandgap from 3.4 eV (GaN) to 6.2 eV (AlN) and remarkable material characteristics. However, achieving efficient *p*-type doping in high aluminum composition (Al%) AlGaN remains a formidable challenge. This study presents an alternative approach to address this issue by fabricating a $p^+$Si/$n^-$AlN/$n^+$AlGaN heterojunction structure by following the semiconductor grafting technique. Atomic force microscopy (AFM) analysis revealed that the AlN and the nanomembrane surface exhibited a smooth topography with a roughness of 1.96 nm and 0.545 nm, respectively. High-angle annular dark field scanning transmission electron microscopy (HAADF-STEM) confirmed a sharp and well-defined Si/AlN interface, with minimal defects and strong chemical bonding, crucial for efficient carrier transport. X-ray photoelectron spectroscopy (XPS) measurements demonstrated a type-I heterojunction with a valence band offset ($\Delta E_v$) of 2.73-2.84 eV and a conduction band offset ($\Delta E_c$) of 2.22-2.11 eV. The pn diode devices exhibited a linear current-voltage (I-V) characteristic, an ideality factor of 1.92, and a rectification ratio of $3.3 \times 10^4$, with a turn-on voltage of 3.9 V, indicating effective p-n heterojunction. Temperature-dependent I-V measurements showed stable operation up to 90°C. The heterojunction's high-quality interface and electrical performance showcase its potential for advanced AlGaN-based optoelectronic and electronic devices.

**Keywords:** Si/AlN heterojunction, UWBG materials, Band alignment, Semiconductor grafting, p-type doping

## 1. Introduction

Ultra-wide bandgap (UWBG) semiconductor materials, with bandgaps larger than 3.4 eV and high breakdown electric field, are expected to demonstrate the ability to promote high-power RF electronics and the potential for short-wavelength deep-UV photonics[1, 2]. These capabilities of the UWBG materials could pave the way for new horizons in the fields of electronics and photonics. Among UWBG semiconductors such as $Ga_2O_3$, Diamond, and BN, $Al_xGa_{1-x}N$ stands out as one of the most promising choices [3-7]. AlGaN-based devices are widely used in high-power and high-frequency electronics, optoelectronics, and biosensors due to their high electron mobility, wide bandgap, and thermal stability, making them suitable for demanding applications such as RF amplifiers, and UV LED [8-10]. In addition, AlGaN is a direct bandgap material with a tunable bandgap ranging from 3.4 eV (GaN) to 6.2 eV (AlN) based on the aluminum composition. This tunability makes it ideal for manufacturing light emitters, including LEDs and laser diodes, as well as light-sensing devices such as detectors. [11, 12].

Although AlGaN-based devices have significant potential, existing challenges still halt the progress both in material growth and device fabrication [1, 13]. It is challenging to achieve a high concentration of *p*-type doping in the AlGaN due to the material's high activation energy, the strong self-compensation effect, and the tendency to quickly transform into a deep acceptor [2, 14, 15]. As the aluminum composition (Al%) in AlGaN alloys increases beyond 50%, achieving reliable and efficient *p*-type doping becomes increasingly challenging. While *n*-type AlGaN with aluminum concentrations up to 80% can be readily obtained, the formation of stable and highly conductive *p*-type layers becomes progressively more difficult, especially when the Al% exceeds 50% [16, 17]. The deep acceptor levels in high Al% AlGaN are crucial as they can neutralize intentional dopants and affect the performance of electronic and photonic devices through carrier trapping and non-radiative recombination [18, 19]. To address the high activation energy, different methods have been introduced to prevent the self-compensation effect by reducing the activation energy, and increasing the solubility of Mg. A few of these methods include modulation doping, delta (*δ*) doping, and polarization-induced doping [20-22]. However, the problem of poor conductivity in *p*-AlGaN with a high aluminum composition remains unresolved. Finding an efficient and innovative solution to this issue is crucial for unlocking the full potential of AlGaN-based devices.

A pioneering technique of grafting semiconductors emerges as a potential solution that can create heterostructure among dissimilar materials. An ultrathin dielectric, ultrathin oxide (UO) in the sub-nanometer range is used at the interface for passivation and quantum tunneling. Various heterostructures have been realized with this technique, Si/Ge, Si/GaAs, Si/GaAsSb, GaAs/diamond, GaAs/$Ga_2O_3$, Si/$Ga_2O_3$, Si/GaN, and GaAs/GaN, etc. [23-29], which could have otherwise been difficult, or not impossible, with the current material growth techniques due to their lattice and structural mismatch. Attempting this technique, in this study, we demonstrated the fabrication of a $p^+$Si/$SiO_2$/$n^-$AlN/$n^+$AlGaN heterojunction structure, addressing the challenges associated with *p*-type doping in high aluminum content (Al%) AlGaN alloys. A 0.5 nm $SiO_2$ UO layer was introduced in between Si and an AlN/AlGaN epi layer. Comprehensive characterization was performed using X-ray diffraction (XRD), atomic force microscopy (AFM), scanning electron microscopy (SEM) and high-angle annular dark field scanning transmission electron microscopy (HAADF-STEM) to analyze the structural, surface and interface properties. A band alignment study was also conducted using the X-ray photoelectron spectroscopy (XPS) core-level method to evaluate the interface properties. Additionally, I-V measurements at room and elevated temperatures were performed to assess the electrical properties. This $p^+$Si/$n^-$AlN/$n^+$AlGaN heterojunction undergoes thorough characterizations to understand the properties of the heterojunction produced. By integrating a *p*-Si NM with the AlN/AlGaN epitaxial layers, this study provides the missing *p*-type doping in high Al% AlGaN, with the potential to enable the development of advanced optoelectronic and electronic devices that leverage the unique properties of these UWBG materials.

## 2. Experiments and Methods

*Materials growth*

Single crystalline AlN (30 nm), AlGaN (300 nm), and AlN (100 nm) epilayers were grown on a 430 µm c-plane sapphire substrate using a molecular beam epitaxy (MBE) system under ultra-high vacuum conditions. The initial AlN layer, deposited directly on the sapphire substrate, served as a nucleation layer. Subsequently, a 300 nm Si-doped $n^+$AlGaN layer was grown, with the Si cell temperature maintained at 1250 °C. Hall effect measurements indicated an electron concentration of approximately $5 \times 10^{18}$ cm$^{-3}$. The Al content in the AlGaN layer ranged between 70-75%. The quality of the substrate and detailed analysis of the epilayers are further discussed in the results section.

A 6-inch Soitec® *p*-type silicon (100)-on-insulator (SOI) substrate, initially with a 205 nm Si device layer and a 400 nm buried oxide (BOX) layer, underwent a 1050 °C thermal oxidation for 12 min, resulting in a 36 nm thick screen oxide. Room temperature boron implantation ($3 \times 10^{15}$ cm$^2$, 15 keV, 7° angle) was followed by a 950 °C, 40 min thermal activation process. The final Si device layer had a thickness of 180-185 nm, with boron doping concentrations ranging from $9.7 \times 10^{19}$ cm$^{-3}$ (top) to $9.5 \times 10^{19}$ cm$^{-3}$ (bottom). The top screen oxide measured ~52 nm. Silvaco Athena® simulations guided the process, ensuring the maintained monocrystalline properties of the $p^+$ Si device layer post-dopant activation [30].

*Synthesis of nanomembrane*

The process flow for the device fabrication is shown in **Figure 1**. To fabricate the $p^+$Si/$n^+$AlGaN diode, the *p*-type silicon (*p*-Si) SOI wafer was diced into 4×4 mm$^2$ pieces using a Disco DAD3221 Automated wafer saw, **Figure 1a**. To release the nanomembranes (NMs) of *p*-Si from the epitaxial wafer, the diced pieces underwent a standard semiconductor cleaning process. This involved sequential ultrasonic baths in acetone, isopropyl alcohol (IPA), and deionized (DI) water, each lasting 10 minutes, followed by drying with dry nitrogen (N$_2$). Next, a photoresist (PR) 1813 was applied to the samples and 9×9 µm$^2$ holes were patterned using UV-photolithography. The reactive ion etcher (RIE) (Plasma-Therm 790) was employed to etch through the patterns down to the sacrificial layer of SiO$_2$, using an RF power of 100W and a gas combination of O$_2$ and SF$_6$. The PR was removed using acetone and IPA. Subsequently, the samples were immersed in a 49% HF solution for 3 hours to undercut the SiO$_2$ sacrificial layer, **Figure 1b**. After completing the undercut, any remaining HF residues were removed with DI water, and the samples were dried with N$_2$. The *p*-Si nanomembrane was then picked up using a polydimethylsiloxane (PDMS) stamp, **Figure 1c**. The AlN/AlGaN substrate was cleaned with similar standard cleaning procedures, followed by the deposition of a 0.5 nm layer of SiO$_2$ using a glove-box integrated atomic layer deposition (ALD) (Ultratech/Cambridge Nanotech S100). The SiO$_2$ was deposited by ALD because of the conformal coverage it provides over the substrate even at sub- nanometer thicknesses. The released Si NM was then transferred to the AlN/AlGaN substrate after the removal of PDMS, Figure 1d. The transferred Si NM was weakly bonded to the substrate through Van der Waals force, so it underwent rapid thermal annealing (RTA) at an optimized temperature of 350 °C for 5 minutes to form robust chemical bonding. The detailed schematic process of NM transfer is been shown in our previous work [29].

*Formation of the $p^+$Si/nAlN/$n^+$AlGaN Heterojunction Diode*

Metal deposition was carried out using an electron beam evaporation (Angstrom Engineering Nexdep Physical Vapor Deposition Platform, e-beam) tool to create metal pads for both the anode and cathode. The process began with anode metal deposition, the anode pattern was defined through lithography, and a metal stack of Ni/Au with a thickness of 10/100 nm was deposited using the same e-beam machine, **Figure 1e**. Followed by the cathode mesa patterning through photolithography. Subsequently, selective etching was performed using an RIE (Plasma-Therm, USA) tool, using RF power of 100 W with $O_2$ gas for Si and RF power of 100 W with $Cl_2$ and Ar for AlGaN, to etch down through the *p*-Si NM and AlN to reach the AlGaN cathode region. The trench depth (~280 nm) of the resulting structure was confirmed using an optical profilometer from KLA USA, **Figure 1f**. Next, the cathode pattern was defined using photolithography, and a stack of metals (Ti/Al/Ni/Au: 20/100/50/100 nm) was deposited at a constant rate of 0.5 Å per second using the e-beam technique with a power of 500 W. The base pressure during this deposition was maintained at $5\times10^{-7}$ torr, **Figure 1h**. Following the metal deposition for both the cathode and anode, individual diodes were isolated by lithography patterning and RIE etching, employing $O_2$ gas and an RF power of 100 W for 2 minutes. The final device structure is shown in **Figure 1h**.

*Characterizations*

To characterize the fabricated heterojunction and devices, various analytical techniques were employed. For structural analysis of the epitaxial layers shown in **Figure 2a**, high-resolution X-ray diffraction (HR-XRD) was performed using a Malvern Panalytical Empyrean diffractometer with a Cu-Kα source in 2θ-ω geometry, as shown in **Figures 2b and 2c**. A brightfield optical microscope image of the nanomembranes (NM) was obtained, displayed in **Figure 2d**. Microstructural analysis of the devices was conducted using a Zeiss LEO 1530 Field-Emission Scanning Electron Microscope (FESEM) in planar geometry, confirming device uniformity and consistency, **Figure 2e**. Elemental composition at the interface was analyzed via energy dispersive spectroscopy (EDS) using a Zeiss Gemini 300 SEM, as shown in **Figure 2f**.

The cross-sectional analysis was done using the scanning tunneling electron microscopy (STEM). Cross-sectional TEM specimen was prepared by an in-situ FIB lift out method. A Thermo-Fisher Helios 650 Xe plasma dual beam FIB instrument was used. A 30 keV ion beam was used for thin lamella cutting and initial thinning and a 5 keV ion beam used for final thinning and polishing. A Thermo-Fisher Talos 200F which was operated at 200keV was used for imaging and diffraction and element mapping, **Figure 3**.

Atomic Force Microscopy (AFM) analysis was conducted using the Bruker Dimension Icon AFM. AFM provides valuable insights into the surface morphology and roughness of the nanomembranes, shown in **Figure 4**. Kelvin probe force microscopy (KPFM) was employed to assess the work function difference at the *Si/AlN* boundary, using a platinum/iridium-coated conductive probe. This probe, with a resonance frequency of ~75 kHz, a ~25 nm, and a spring constant of about 3 N.m$^{-1}$, facilitated precise measurements at the *p-n* junction.

In **Figure 5**. the Si/AlN heterostructure's band alignment was studied using a Thermo Scientific K Alpha X-ray Photoelectron Spectrometer with an Al Kα source (hν = 1486.68 eV) in a $2 \times 10^{-8}$ Torr high vacuum system. Measurements were made with a pass energy of 50 eV, a spot size of 400 µm, dwell times at 0.5 s, and a step size of 0.05 eV. To evaluate the electrical performance of the devices, current-voltage (I-V) measurements were carried out using a Keithley 4200 Parameter Analyzer, **Figure 6.**

## 3. Results and Discussion

The crystalline and structural properties were analyzed using the HR-XRD measurements as displayed in **Figure 2b**. A 2theta-omega scan was conducted for the bare substrate with AlN/AlGaN/AlN layers and an AlN (0002) peak is observed at 35.97°, **Figure 2b**. The inset image shows the log scale and a more prominent AlN peak can be observed. In addition, there is a small AlGaN (0002) peak around 35.5°. The peaks are in close accordance with previous studies [31, 32]. Along with AlGaN/AlN, there is a sharp substrate peak at 41.8° for sapphire ($Al_2O_3$) substrate.

Further, the full width at half maximum (FWHM) was evaluated for the AlN/AlGaN epi layer as displayed in **Figure 2c**. The FWHM of the XRD 2theta-omega scan of the AlGaN (0002) peak is 0.16°. The FWHM values are indicative of the dislocation density within the grown layer, which ultimately influences the performance of the optoelectronic device developed from these layers [33]. The results suggest a considerable crystalline quality of the AlN/AlGaN substrate.

In addition, the plain view of the NM was observed using bright optical microscope images, where a clean NM surface and holes were observed, as displayed in **Figure 2d.** The NM holes were 9×9 $\mu m^2$ wide with a spacing of 50 µm in between the holes. The dark field microscope image was taken which shows no irregularity or non-uniformity on the NM surface during the transfer process, **Figure S1a**. **Figure S1b** displays the whole sample image with an area of 2.8×2.8 $mm^2$, with high yield, nanomembrane integrity and consistency. The enlarged few devices image is shown in **Figure 2e** on a 50 µm scale bar, a high device uniformity over the substrate can be observed. The top anode diamond structure and the bottom cathode structure are quite uniform as displayed.

The layer structure was analyzed using the energy dispersive spectroscopy on the scanning electron microscope (EDS-SEM) technique displayed in **Figure 2f**, at the epi substrate/*p*-Si lateral interface where high Al counts were observed at the substrate side and increased Si counts after the interface. An inset SEM image on top shows the substrate and Si NM interface. The transfer of membranes to create a heterojunction was a critical step, achieved with high yield and surface smoothness using polydimethylsiloxane (PDMS). To further analyze the Si/AlN heterojunction tunneling electron microscopy was done.

The structural analysis of the Si/AlN interface was performed using high-angle annular dark field scanning transmission electron microscopy (HAADF-STEM). **Figure 3a** presents a cross-sectional view of the device, showing four well defined layers: Si, AlN, AlGaN and a substrate. A closer examination of the Si/AlN interface is provided in **Figure 3b**, highlighting the two layers with a clear interface. The post grafting TEM images in **Figure 3c** reveal a sharp, well-defined interface with excellent crystalline quality, confirming the successful transfer and strong chemical bonding of the Si NM to the AlN, without any visible air gaps or defects such as misfit dislocations. The presence of some amorphous regions along with a thin $SiO_2$ layer has been observed at the junction. The total thickness of the interface layer is around 5 nm, which is wider than the as-deposited ALD-$SiO_2$ layer (~0.5 nm), likely due to oxygen atom redistribution and chemical bonding with surrounding materials during thermal annealing. An optimized interface layer is crucial for improving carrier collection, as a thinner oxide layer enhances the likelihood of carrier tunneling, leading to higher device efficiency. Nonetheless, carriers can still be transported via quantum tunneling. Additionally, an atomic percentage versus position graph has been extracted from the cross-sectional analysis, illustrating the variation in elemental concentrations across different regions, as shown in **Figure S2**.

The transferred nanomembrane (NM) surfaces were meticulously analyzed for quality using atomic force microscopy (AFM) and optical microscopy, as depicted in **Figure 4**. **Figure 4a** shows the AlN surface roughness measurement for the epi structure, which was calculated to be 1.9 nm, where a high-quality surface morphology is observed with atomic steps as discussed by Cheng et al. [34]. Similarly, the transferred NM surface roughness, **Figure 4b,** was calculated to be 0.545 nm showing a smooth surface, hence creating an even interface. To better observe the lateral boundary and measure the NM thickness an AFM technique was used, as evident from **Figure 4c**, a smooth top view interface can be seen between the substrate and the *p*-Si NM, a height difference was observed, and the NM was found to be ~185 nm thick, as shown in **Figure 4d.**

We utilized scanning Kelvin probe force microscopy (KPFM) for detailed analysis of the junction barrier and surface potential distribution at the Si/AlN boundary, enabling us to capture crucial junction characteristics such as built-in potential and depletion layer widths [35]. 3D KPFM image revealed a notable interface potential difference within the Si/AlN structures **Figure 4e**. KPFM provided a visual representation of surface potential variation, highlighting the potential drop across the heterostructures. The observed average surface potential difference between Si and AlN layers, approximately 108 mV as indicated in **Figure 4f**, reveals that Si Fermi level is 108 meV higher than that of AlN, with a depletion layer width of ~0.45 µm. The electric field within the depletion region can be determined using the built-in potential and the depletion layer width, yielding a value of $2.4 \times 10^5$ V·m$^{-1}$. This observation underlines the presence of charge transfer and the establishment of a built-in potential at the heterojunction's interlayer, serving as the foundational mechanism for the operation of rectification heterojunction devices.

To gain insights into the band offset of the Si/AlN heterojunction three distinct samples were prepared: pristine Si, pristine AlN, and the grafted Si/AlN heterostructure. The samples were analyzed using Kraut et al.'s approach [36, 37]. This technique assesses the band alignment at thin film heterostructure interfaces by comparing the core level (CL) and valence band maximum (VBM) binding energies of each heterostructure component. Following that, the difference in CL energies across the heterostructure enables the valence band offset ($\Delta E_v$) calculation at the interface between Si and AlN using the following equation.

$$\Delta E_v = \left(E_{Si\ 2p}^{Si} - E_{VBM}^{Si}\right) - \left(E_{Al\ 2p}^{AlN} - E_{VBM}^{AlN}\right) + \left(E_{Al\ 2p}^{AlN} - E_{Si\ 2p}^{Si}\right)_{interface} \quad (1)$$

Where $E_{Si\ 2p}^{Si}$ and $E_{Al\ 2p}^{AlN}$ represent the Si 2p and Al 2p binding energy position of pristine Si and AlN film, while E$_{VBM}$ denotes the valence band maximum (VBM) position of respective samples. $E_{Al\ 2p}^{AlN} - E_{Si\ 2p}^{Si}$ represents the energy difference between the Al 2p and Si 2p of the Si/AlN heterostructure at the interface. The valence band maximum (VBM) energy was identified by extending the straight part of the valence band's low-energy edge to intersect the spectral baseline [38].

As shown in **Figure 5**, $E_{Si\,2p}^{Si}$ and $E_{Al\,2p}^{AlN}$ exhibit core level peak positions at 98.91 and 73.32 eV, respectively. The VBM values of pristine Si and AlN are measured to be 0.04 and 2.49 eV. **Figures 5e and 5f** display the Al 2p and Si 2p core levels with peak values of 74.26 and 99.46 eV, respectively, from the grafted Si/AlN heterostructure. By applying the derived experimental data to equation (1), the valence band offset (VBO) for the Si/AlN heterostructure was calculated to be 2.84±0.2 eV.

To calculate the conduction band offset ($\Delta E_c$) for the Si/AlN heterostructure was derived using the bandgap values of Si (1.12 eV) and AlN (6.07 eV) (also verified with photoluminescence for this epi sample) [39], along with the previously determined valence band offset ($\Delta E_v$), are substituted in equation 2. This calculation yielded a $\Delta E_c$ value of 2.11±0.2 eV. The overall band alignment diagram is shown in **Figure 5 (g)**. The Si/AlN heterojunction is identified as a Type I heterojunction.

$$\Delta E_c = E_g^{Si} + \Delta E_V - E_g^{AlN} \qquad (2)$$

In Figure S6, we also showed the XPS results of $E_{N\,1s}^{AlN}$ and calculated the band offset values. The valence band and conduction band offsets show 2.73 ±0.2 eV and 2.22±0.2 eV, which are highly consistent with the values calculated using $E_{Al\,2p}^{AlN}$.

The electron affinity (χ) of AlN calculated in this work using Si (χ = 4.05 eV) was approximately 1.83-1.94 eV. The results are close to the Si/AlN heterojunction by epitaxy growth [40]. Various values for the electron affinity of AlN have been reported, reflecting different measurement techniques, theoretical approaches and also possibly the different AlN qualities. For instance, King et al. obtained an electron affinity of AlN around 2.45 eV using Si as a reference [40]. In contrast, Grabowski et al. measured a significantly lower electron affinity of about 0.4 eV using Ultraviolet Photoelectron Spectroscopy (UPS) [41]. Utilizing density functional theory calculations, Tsai et al. reported an electron affinity of approximately 1.4 eV [42]. Additionally, Bermudez et al. reported an electron affinity of around 0.6 eV, derived from calculations based on the electron affinity of $Al_xGa_{1-x}N$ [43]. These discrepancies highlight the variability in reported values and the influence of different methodologies on the determination of electron affinity for AlN. Further studies are required to achieve a more accurate and consistent understanding of the electron affinity of AlN. Due to the uncertainty values of the AlN electron affinity, it is still premature to deduce that the Si/AlN/AlGaN heterojunction formed in this study has been a successfully grafted one. As shown in our recent studies, a successful grafting would result in a band alignment that follows the electron affinity rule [27, 29]

The Si/AlN/AlGaN pin diode performance for a 45x45 µm² anode area device is shown in **Figure 6**, a linear scale I-V plot is shown in **Figure 6a,** and an inset image of the device structure is employed for the measurement of the device. Similarly, a log scale I-V is plotted in **Figure 6b,** and the ideality factor (IF) and the rectification ratio were calculated to be 1.92 and $3.3 \times 10^4$ at ±4 V. The shift of the minimum current point to a positive bias (~1.2 V) instead of zero voltage can be attributed to slow charging and discharging process originating from the high equivalent resistance under reverse bias (i.e., the reverse bias current is too small). The high ideality factor can be attributed to the non-ideal Si/AlN interface conditions, i.e., the surface Fermi-level pinning on the AlN, which necessitates an additional detailed study. Additionally, the effects of series resistance, like poor ohmic contact resistance on the $n^+$ AlGaN layer may also contribute to this increase. Nevertheless, it can be seen from the linear scale plot that the diode has a typical rectifying characteristic and a turn-on voltage is about 3.9 V obtained from linear extrapolation [44]. At +5 V device showed a current density and a specific on-resistance of 49 mA/cm$^{-2}$ and 101.3 Ω. cm$^2$, respectively. The device performance can be further enhanced by optimizing the contact resistance through a more appropriate thermal budget and by applying efficient passivation layers.

Furthermore, multiple measurements were conducted to check the device consistency as shown in **Figure S3**. For the single device I-V plotted in **Figure 5,** five measurements were taken on the same device to check the stability, indicating consistent performance as displayed in **Figures S3a and S2b**. In addition, multiple devices that are randomly selected were measured to check the uniformity across the substrate and all five measured devices showed regularity as shown in **Figures S3c and S3d**.

**Figure 6c** highlights the breakdown characteristics observed in some devices chosen at random noting a breakdown voltage between 17-19 V. This variability is attributed to the inherently random nature of the avalanche phenomenon. The breakdown was catastrophic, destroying the device permanently as highlighted in the optical microscopic image of **Figure 6d**. The devices that are measured for breakdown measurements showed a similar trend. Since the *i*-AlN layer sustains the reverse bias, the breakdown electrical field of the 30 nm AlN is estimated to be 5.6-6.3 MV/cm. It is noted that no field plate was employed in the simple diode structure.

Temperature-dependent forward bias I–V characteristics of the $p^+$Si/$n^+$AlGaN heterojunction were also measured for a randomly selected device with the temperature being varied from room temperature (RT) to 90°, **Figure S4a**. From the forward bias measurements, the dopant activation energy in AlN was approximately 196 meV as shown in **Figure S4b**. The results obtained in this study align well with recent studies, which report AlN's dopant activation energies in the range of approximately 200 meV to 300 meV [45-47]. It is important to note, however, that in AlN, the activation energy is highly dependent on the growth conditions, as detailed in the recent findings by Hermann et al [48].

Additionally, reverse bias measurements were conducted at various temperatures, as shown in **Figure S5a.** From these measurements, the defect activation energy in AlN was estimated to be 106 meV at -8 V, illustrated in **Figure S5b**. Notably, when the reverse bias was increased to -12 V, the defect activation energy decreased, calculated to be 71 meV, as presented in **Figure S5c.** This potentially implies that the defect energy levels at different depletion regions vary.

## Conclusion

This study demonstrates a fabrication of a $p^+$Si/SiO$_2$/$n^-$AlN/$n^+$AlGaN heterojunction by following a semiconductor grafting method, addressing challenges in $p$-type doping for high aluminum composition AlGaN alloys. HR-XRD confirmed the high crystalline quality of the substrate. AFM showed a smooth AlN surface with a roughness of 1.9 nm and a smoother transferred NM surface at 0.545 nm. High-angle annular dark field scanning transmission electron microscopy (HAADF-STEM) confirmed the presence of a sharp, well-defined Si/AlN interface with outstanding crystalline quality and minimal defects. In addition, band alignment studies using XPS measurements revealed a type-I heterojunction with a $\Delta E_v$ of 2.84±0.2 eV and a $\Delta E_c$ of 2.11±0.2 eV. The diodes exhibited an ideality factor of 1.92, a rectification ratio of $3.3 \times 10^4$, and a turn-on voltage of 3.9 V. The demonstrations showed the promise of heterogeneous integration in overcoming the limitations of high Al% AlGaN, offering a possible way for advanced optoelectronic and electronic devices that leverage the unique properties of AlGaN-based materials. Further detailed study on the interface properties of Si/AlN formed by a similar approach will be needed.

**ACKNOWLEDGMENTS**

The device fabrication was supported by Defense Advanced Research Projects Agency (DARPA) under the H2 program (HR0011-21-9-0107). The authors acknowledge Prof. Ping Wang for his help with the growth of the AlN/AlGaN heterostructures.

**DATA AVAILABILITY**

The data that support the findings of this study are available from the corresponding author upon reasonable request.

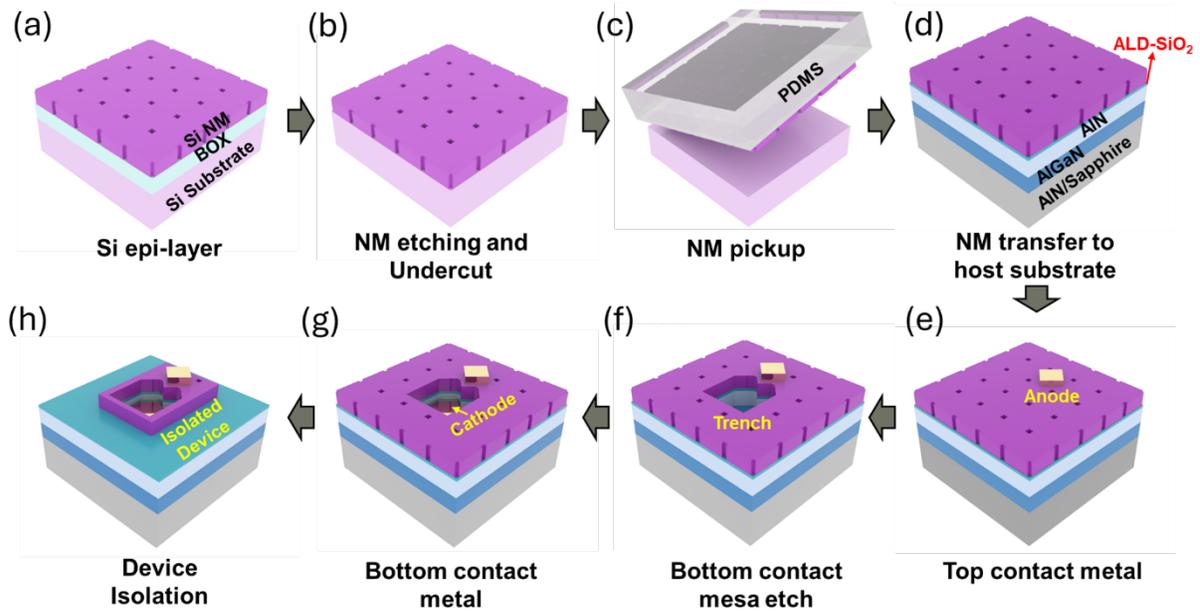

**FIG. 1.** (a) Process flow illustration of the Si/AlN heterojunction and *p-n* diodes. (a) Si substrate cleaning and nanomembrane (NM) holes pattering. (b) Etching and undercutting the sacrificial SiO$_2$ layer. (c) NM was picked up using the polydimethylsiloxane (PDMS) stamp. (d) AlN/AlGaN substrate was cleaned and coated with the ALD-SiO$_2$ (~ 0.5 nm). (e) Top metal pattering and deposition with Ni/Au metal stack. (f) Defining Mesa, plasma etching of the Si NM and AlN to reach the highly doped AlGaN layer. (g) The bottom metal contact lithography and metal deposition with Ti/Al/Ni/Au metal stack. (h) Device isolation and the final device structure.

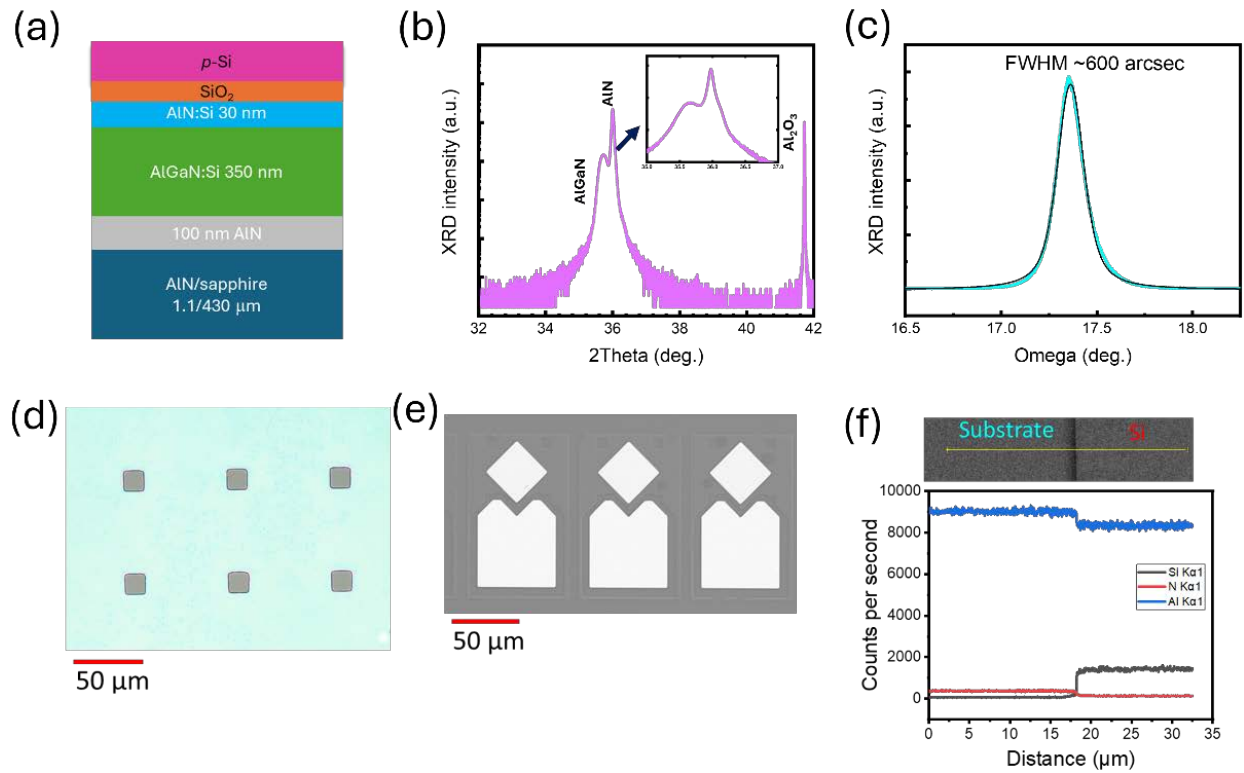

**FIG. 2.** (a) Epi layer used for the device fabrication. Structural and Crystallographic properties XRD 2theta-omega scan (b) On the AlN/AlGaN/AlN/Sapphire substrate (c) FWHM of the AlGaN (0002) and the fitting curve with an FWHM value of 0.61°. (d) Brightfield Optical microscopic image of the NM, where the 9 × 9 µm$^2$ etch holes with 50 µm are apparent. (e) Scanning electron microscopy (SEM) device structure image showing uniform device structures. (f) Energy dispersive spectroscopy on the scanning electron microscope (EDS-SEM) was used to check the presence of different elements at the interface, displaying the existence of NM on the substrate.

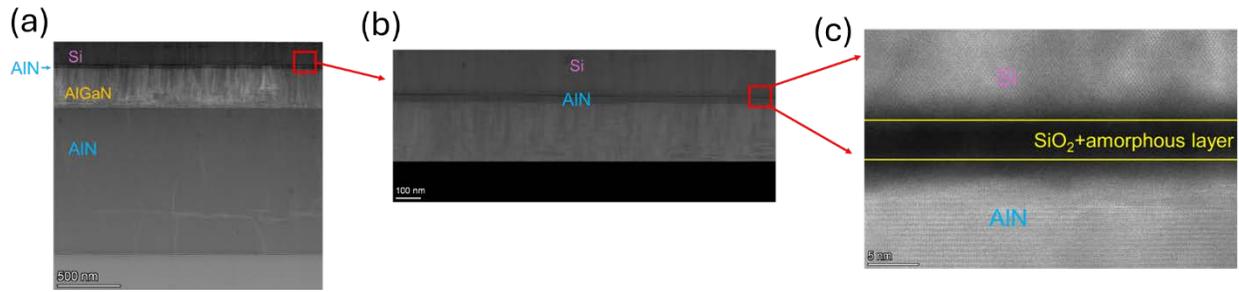

**FIG. 3.** High-Angle Annular Dark Field Scanning Transmission Electron Microscopy (HAADF-STEM) of the grafted Si/AlN heterojunction. (a) The cross-sectional image of the Si NM grafted onto the AlN substrate. (b) STEM image of the interface. (c) An enlarged image of the Si-AlN interface with the amorphous layer.

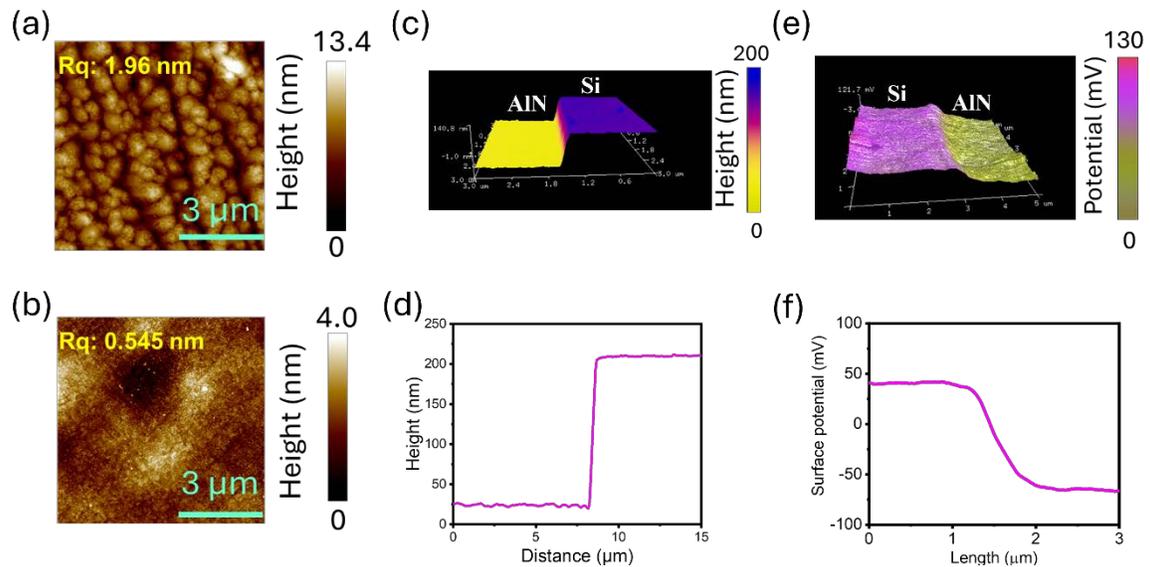

**FIG. 4.** Surface morphology analysis of the sample. (a) The Atomic force microscopic image of the substrate (b) AFM image of the nanomembrane. (a) Atomic force microscopic image at the substrate and Si NM interface showing a smooth interface and the scale bar is 5 µm (d) The height difference profile observed at the interface and the NM thickness of ~185 nm. Kelvin Probe Force Microscopy study (a) 3D profile image and (b) the potential difference on either side of the heterojunction with Si having higher potential than the AlN.

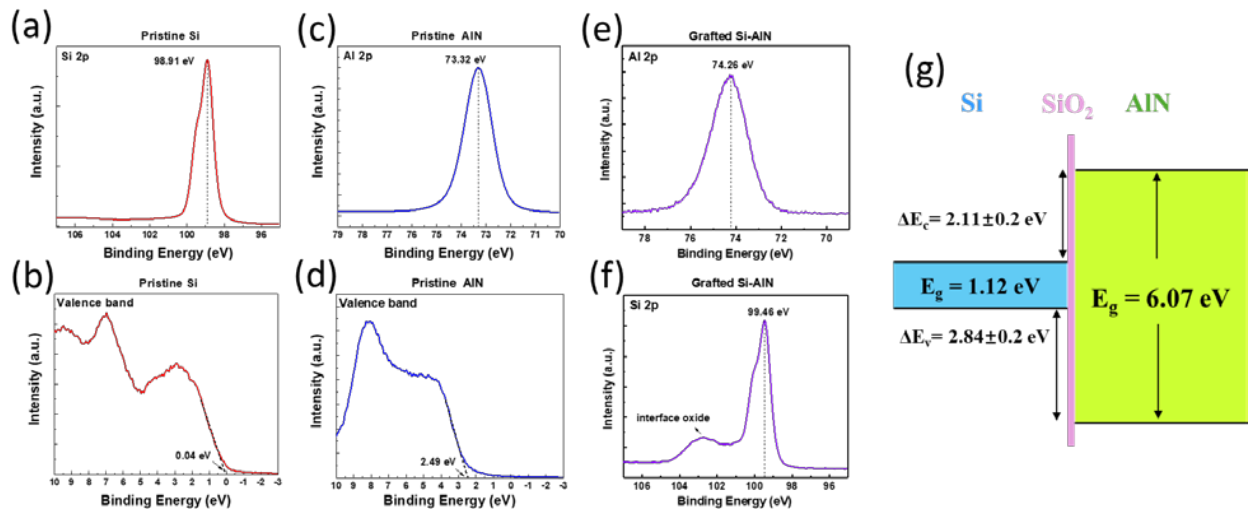

**FIG. 5.** XPS core level spectra of (a) Si 2p and (b) valence band spectra of the pristine Si; (c) core level spectra of Al 2p and (d) valence band spectra of the pristine AlN; core level spectra of (e) Al 2p and (f) Si 2p of the grafted Si/AlN interface. (g) The band alignment diagram of the Si/AlN heterojunction.

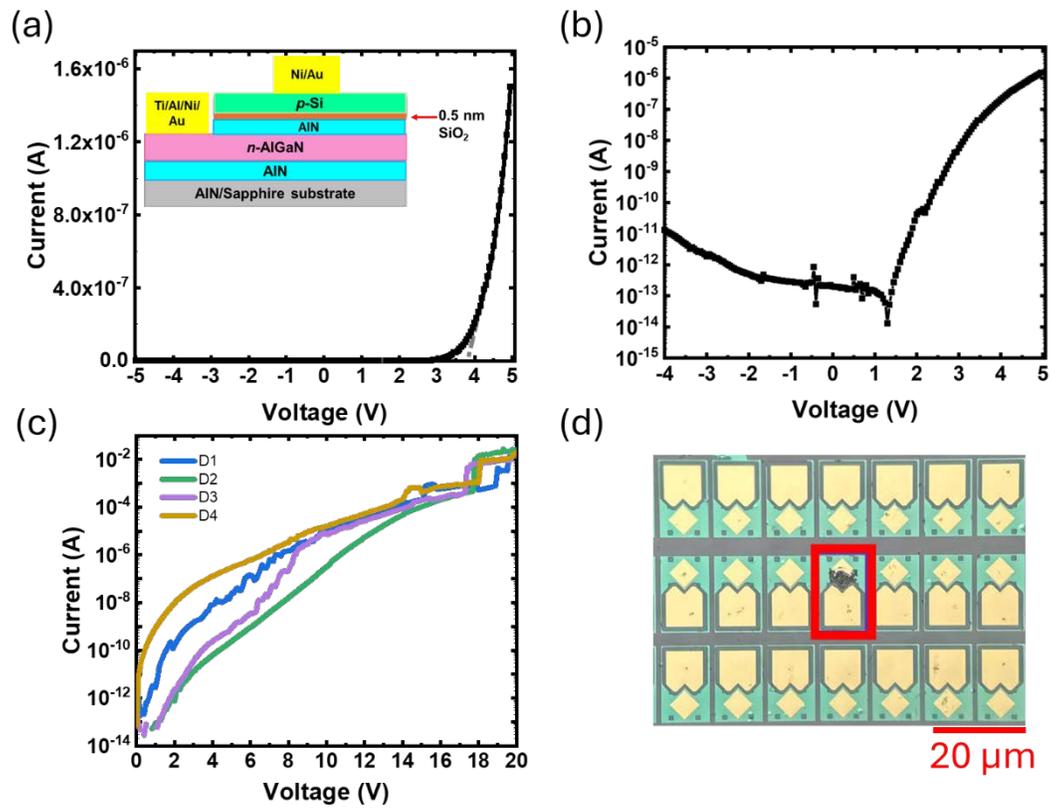

**FIG. 6.** The I-V performance of the device. (a) linear scale with an inset image of the device structure and (b) logarithmic scale. (c) Reverse breakdown characteristics of 4 randomly selected devices. (d) The optical image of multiple devices and a single device in a square with a breakdown effect.

# Supplemental Data

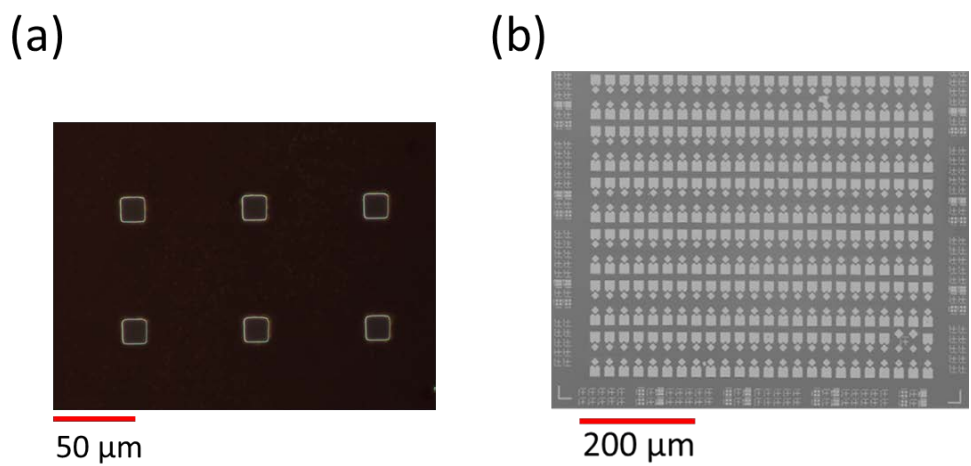

**FIG. S1**. (a) Optical microscopic dark field image of the nanomembrane. (b) Scanning electron microscopy image of the whole device showing uniformity and high yield.

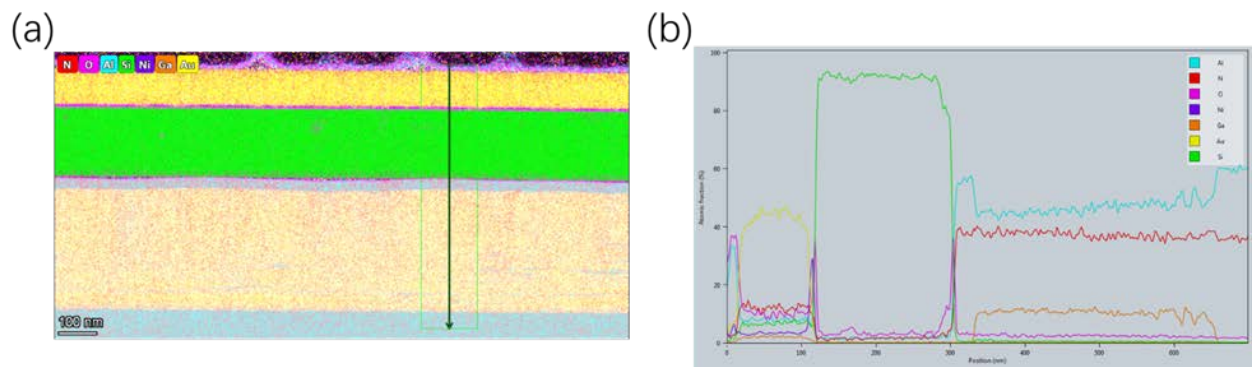

**FIG. S2.** EELS elemental mapping of the Si/AlN heterojunction device. (a) The scanning region for EELS mapping, starting from the top metal layer and progressing through the metal/Si NM interface, Si NM, $SiO_2$, AlN, AlGaN, and down to the substrate. (b) Atomic percentage profile as a function of position (nm) measured across the entire device structure.

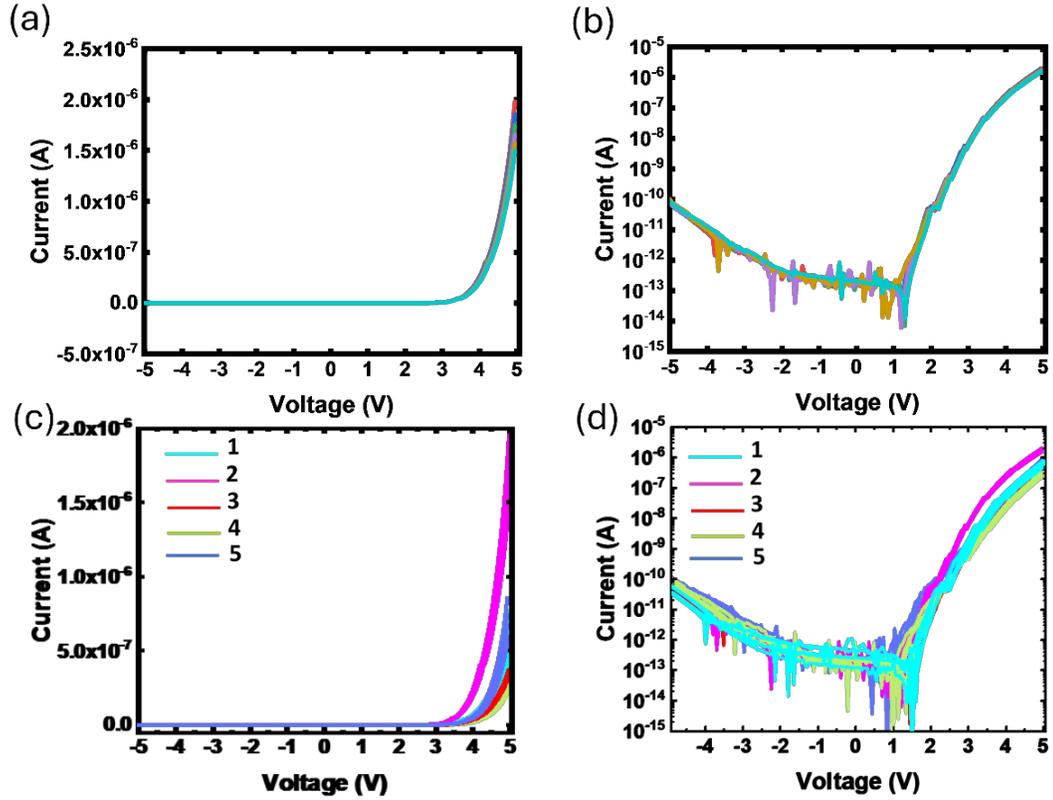

**FIG. S3.** Device stability measurements were conducted, showing: (a) five measurements of a single device displayed on a linear scale, and (b) the same data on a logarithmic scale. (c) five different devices on the wafer were measured and presented on a linear scale, and (d) the corresponding data on a logarithmic scale.

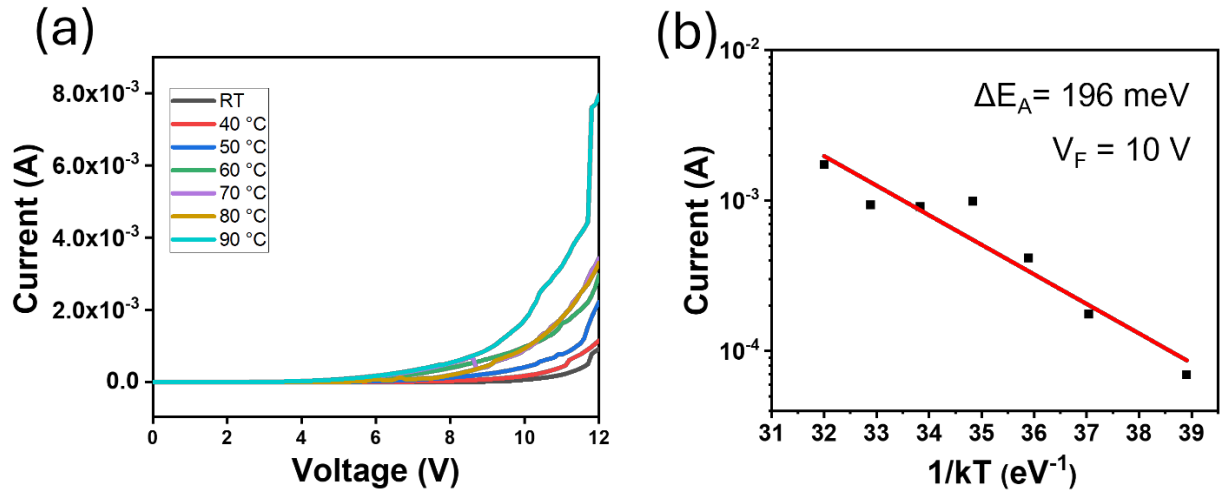

**FIG. S4.** Temperature-dependent I–V characteristics of the $p$-Si/AlN/$n$-AlGaN heterojunction. (a) Forward bias measurement from 0 to +12 V. (b) Current at 10 V forward bias vs $1/kT$. The slope gives an activation energy of around 196 meV.

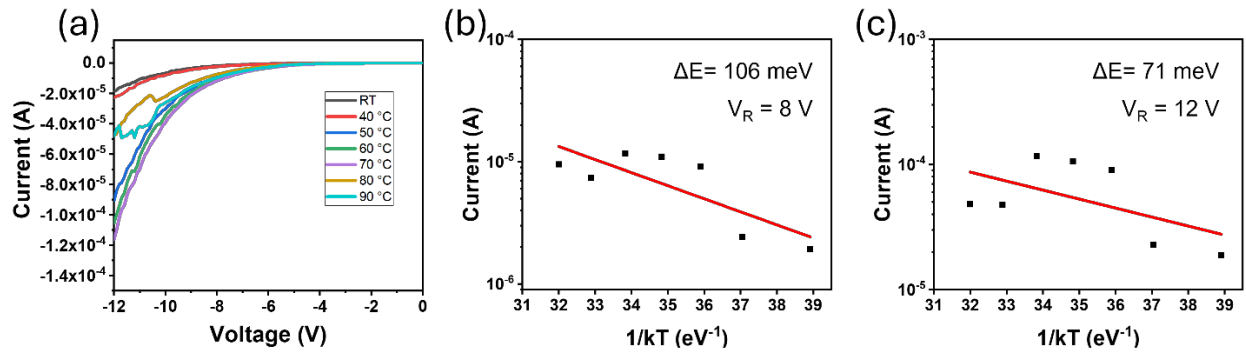

**FIG. S5.** Temperature-dependent reverse bias I–V characteristics of the p-Si/AlN/n-AlGaN heterojunction. (a) Reverse bias measurements 0 to -12 V. (b) Current at -8 V reverse bias vs 1/kT. The slope gives a defect activation energy of around 106 meV. (c) Current at -12 V reverse bias vs 1/kT with an activation energy of 71 meV.

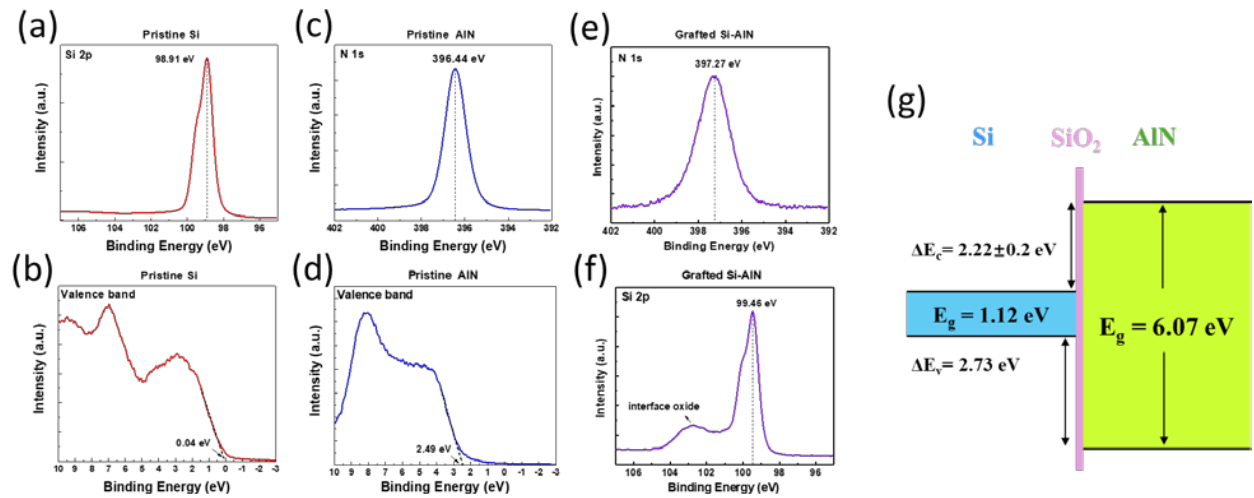

**FIG. S6.** XPS core level spectra of (a) Si 2p and (b) valence band spectra of the pristine Si; (c) core level spectra of N 1s and (d) valence band spectra of the pristine AlN; core level spectra of (e) N 1s and (f) Si 2p of the grafted Si/AlN interface. (g) The band alignment diagram of the Si/AlN heterojunction.